\documentclass{revtex4}
\usepackage{amsfonts}
\usepackage{amsmath}

\input{tcilatex}

\begin{document}

\preprint{}
\title{'Square Root' of the Maxwell Lagrangian versus confinement in general
relativity}
\author{S. Habib Mazharimousavi}
\email{habib.mazhari@emu.edu.tr}
\author{M. Halilsoy}
\email{mustafa.halilsoy@emu.edu.tr}
\affiliation{Department of Physics, Eastern Mediterranean University, G. Magusa, north
Cyprus, Mersin 10 - Turkey.}
\pacs{PACS number}

\begin{abstract}
We employ the 'square root' of the Maxwell Lagrangian (i.e. $\sqrt{F_{\mu
\nu }F^{\mu \nu }}$), coupled with gravity to search for the possible linear
potentials which are believed to play role in confinement. It is found that
in the presence of magnetic charge no confining potential exists in such a
model. Confining field solutions are found for radial geodesics in pure
electrically charged Nariai- Bertotti-Robinson (NBR)-type spacetime with
constant scalar curvature. Recently, Guendelman, Kaganovich, Nissimov and
Pacheva, [Phys.Lett.B704(2011)230] have shown that superposed square root
with standard Maxwell Lagrangians yields confining potentials in spherically
symmetric spacetimes with new generalized Reissner-Nordstr\"{o}m-de Sitter /
-anti-de Sitter black hole solutions. In NBR spacetimes we show that
confining potentials exist even when the standard Maxwell Lagrangian is
relaxed.
\end{abstract}

\maketitle

A power-law extension of the Maxwell action coupled with gravity was
considered by \cite{1,2,3}%
\begin{equation}
I=\frac{1}{2}\int dx^{4}\sqrt{-g}\left( R-2\Lambda -\alpha \mathcal{F}%
^{s}\right) ,
\end{equation}%
in which $s$ and $\alpha $ are real constants, $\mathcal{F}=F_{\mu \nu
}F^{\mu \nu }$ is the Maxwell invariant with $F_{\mu \nu }=\partial _{\mu
}A_{\nu }-\partial _{\nu }A_{\mu }$ and $\Lambda $ stands for the
cosmological constant. The first study with this form of nonlinear
electrodynamic (NED) was made in spherical symmetry and ever since many
authors have considered different aspects / applications of this action \cite%
{2}. Although the original paper \cite{3} considered a conformally invariant
action (i.e. $s=d/4$) this requirement was subsequently relaxed. It was
shown that $s=1/2$ raised problems in connection with the energy conditions 
\cite{4} and for this reason it was abandoned. Nielsen and Olesen \cite{5}
proposed such a magnetic 'square root' Lagrangian (i.e. $\sqrt{F_{\mu \nu
}F^{\mu \nu }}$) in string theory while 't Hooft \cite{6} highlighted a
linear potential term to be effective toward confinement. More recently
Guendelman et. al. \cite{7} investigated confining electric potentials in
black hole spacetimes in the presence of the standard Maxwell Lagrangian.

In this note we suppress the standard Maxwell Lagrangian, keeping only the
'square root' of the Maxwell Lagrangian, to search for confining potentials.
It is known that under the scale transformation, i.e. $x_{\mu }\rightarrow
\lambda x_{\mu },$ $A_{\mu }\rightarrow \frac{1}{\lambda }A_{\mu },$ ($%
\lambda $=cons.) in $d=4$ the latter doesn't remain invariant. Even in this
reduced form we prove the existence of such potentials in some spacetimes
identified as the Nariai-Bertotti-Robinson (NBR) type spacetime. Due to the
absence of Maxwell Lagrangian $\sim F_{\mu \nu }F^{\mu \nu },$ however, the
Coulomb potential will be missing in our formalism. We choose the case $s=1/2
$ in $d=4$ with a general line element 
\begin{equation}
ds^{2}=-f\left( r\right) dt^{2}+\frac{dr^{2}}{f\left( r\right) N\left(
r\right) ^{2}}+R\left( r\right) ^{2}\left( d\theta ^{2}+\sin ^{2}\theta
d\varphi ^{2}\right) ,
\end{equation}%
where $f(r),N(r)$ and $R(r)$ are three unknown functions of $r$. Our choice
of Maxwell $2-$form is 
\begin{equation}
F=E\left( r\right) dt\wedge dr+P\sin \left( \theta \right) d\theta \wedge
d\varphi 
\end{equation}%
in which $P$ stands for the magnetic charge constant and $E(r)$ is to be
determined. From the variational principle the nonlinear Maxwell equation
reads%
\begin{equation}
d\left( \frac{^{\star }\mathbf{F}}{\sqrt{\mathcal{F}}}\right) =0,
\end{equation}%
in which $^{\star }\mathbf{F}$ is dual of $\mathbf{F.}$ Using the line
element one finds 
\begin{equation}
^{\star }\mathbf{F=}ENR^{2}\sin \theta d\theta \wedge d\varphi -\frac{P}{%
NR^{2}}dt\wedge dr,
\end{equation}%
and 
\begin{equation}
\mathcal{F}=-2E^{2}N^{2}+\frac{2P^{2}}{R^{4}}.
\end{equation}%
The nonlinear Maxwell equation yields%
\begin{equation}
\frac{ENR^{2}}{\sqrt{-2E^{2}N^{2}+\frac{2P^{2}}{R^{4}}}}=\frac{\beta }{\sqrt{%
2}}
\end{equation}%
where $\beta $ is an integration constant. This equation admits a solution
for the electric field as%
\begin{equation}
E=\frac{P\beta }{NR^{2}\sqrt{R^{4}+\beta ^{2}}},
\end{equation}%
and therefore%
\begin{equation}
\mathcal{F}=\frac{2P^{2}}{R^{4}+\beta ^{2}}.
\end{equation}%
We note here that $\mathcal{F}$ is positive which is needed for our choice
of square root expression. Variation of the action with respect to $g_{\mu
\nu }$ gives Einstein-Maxwell equations%
\begin{equation}
G_{\mu }^{\nu }+\Lambda g_{\mu }^{\nu }=T_{\mu }^{\nu }
\end{equation}%
in which 
\begin{equation}
T_{\ \nu }^{\mu }=-\frac{\alpha }{2}\left( \delta _{\ \nu }^{\mu }\sqrt{%
\mathcal{F}}-\frac{2\left( F_{\nu \lambda }F^{\mu \lambda }\right) }{\sqrt{%
\mathcal{F}}}\right) .
\end{equation}%
Explicitly we find%
\begin{equation}
T_{\ t}^{t}=T_{\ r}^{r}=-\frac{\alpha }{\sqrt{2}}\left( \frac{P\sqrt{%
R^{4}+\beta ^{2}}}{R^{4}}\right) ,
\end{equation}%
and 
\begin{equation}
T_{\ \theta }^{\theta }=T_{\ \phi }^{\phi }=\frac{\alpha }{\sqrt{2}}\frac{%
P\beta ^{2}}{R^{4}\sqrt{R^{4}+\beta ^{2}}}.
\end{equation}%
Having $T_{\ t}^{t}=T_{\ r}^{r}$ means that $G_{\ t}^{t}=G_{\ r}^{r}$ which
leads to $N(r)=C$ and $R(r)=r.$ Note that $C$ is an integration constant
which is set for convenience to, $C=1$. The Einstein equations admit a black
hole solution for the metric function given by 
\begin{equation}
f\left( r\right) =1-\frac{2m}{r}-\frac{\Lambda }{3}r^{2}-\frac{P\alpha }{%
\sqrt{2}r}\int \sqrt{1+\frac{\beta ^{2}}{r^{4}}}dr.
\end{equation}%
Here by using the expansion $\sqrt{1+t}=1+\frac{1}{2}t-\frac{1}{8}t^{2}+%
\mathcal{O}\left( t^{3}\right) $ for $\left\vert t\right\vert <1$ one finds
for large $r$ (i.e. $\frac{r^{4}}{\beta ^{2}}>1$)%
\begin{equation}
f\left( r_{large}\right) =1-\frac{P\alpha }{\sqrt{2}}-\frac{2m}{r}-\frac{%
\Lambda }{3}r^{2}+\frac{P\alpha \beta ^{2}\sqrt{2}}{12r^{4}}+\mathcal{O}%
\left( \frac{1}{r^{8}}\right) ,
\end{equation}%
and for small $r$ (i.e. $\frac{r^{4}}{\beta ^{2}}<1$) we rewrite $\int \sqrt{%
1+\frac{\beta ^{2}}{r^{4}}}dr=\int \frac{\beta }{r^{2}}\sqrt{1+\frac{r^{4}}{%
\beta ^{2}}}dr$ which implies 
\begin{equation}
f\left( r_{small}\right) =1-\frac{2m}{r}+\frac{P\alpha \beta }{\sqrt{2}r^{2}}%
-\left( \frac{\Lambda }{3}+\frac{P\alpha \sqrt{2}}{12\beta }\right) r^{2}+%
\frac{P\alpha \sqrt{2}}{112\beta ^{3}}r^{6}+\mathcal{O}\left( r^{10}\right) ,
\end{equation}%
where $m$ is an integration constant related to mass. The Ricci scalar of
the spacetime is given by%
\begin{equation}
R=2+\frac{4m}{r^{3}}-\frac{2\sqrt{2}\alpha P\sqrt{r^{4}+\beta ^{2}}}{r^{4}}+%
\frac{\sqrt{2}P\alpha }{\sqrt{r^{4}+\beta ^{2}}}-\frac{\sqrt{2}P\alpha }{%
r^{3}}\int \sqrt{1+\frac{\beta ^{2}}{r^{4}}}dr,
\end{equation}%
which at infinity is convergent while at $r=0$ is singular. For a moment in
order to see the structure of the electromagnetic field (3) we resort to the
flat spacetime given by the line element 
\begin{equation}
ds^{2}=-dt^{2}+dr^{2}+r^{2}\left( d\theta ^{2}+\sin ^{2}\theta d\varphi
^{2}\right) .
\end{equation}%
The electric field reads as

\begin{equation}
E=\frac{P}{r^{2}\sqrt{1+\frac{r^{4}}{\beta ^{2}}}}
\end{equation}%
which results in the potential%
\begin{equation}
V=-P\int \frac{dr}{r^{2}\sqrt{1+\frac{r^{4}}{\beta ^{2}}}}.
\end{equation}%
Here also we use the expansion $\frac{1}{\sqrt{1+t}}=1-\frac{1}{2}t+\frac{3}{%
8}t^{2}+\mathcal{O}\left( t^{3}\right) $ for $\left\vert t\right\vert <1$ to
obtain%
\begin{equation}
V\left( r_{small}\right) =-P\int \frac{dr}{r^{2}}\left( 1+\frac{r^{4}}{\beta
^{2}}\right) ^{-\frac{1}{2}}=\frac{P}{r}+\frac{Pr^{3}}{6\beta ^{2}}-\frac{%
3Pr^{7}}{56\beta ^{4}}+\mathcal{O}\left( r^{11}\right) ,
\end{equation}%
for small $r$ and%
\begin{equation}
V\left( r_{large}\right) =-P\beta \int \frac{dr}{r^{4}}\left( 1+\frac{\beta
^{2}}{r^{4}}\right) ^{-\frac{1}{2}}=\frac{P\beta }{3r^{3}}-\frac{P\beta ^{3}%
}{14r^{7}}+\frac{3P\beta ^{5}}{88r^{11}}+\mathcal{O}\left( \frac{1}{r^{15}}%
\right) ,
\end{equation}
for large $r.$ It is readily seen that the magnetic charge $P$ is
indispensable for an electric solution to exist in the flat spacetime.

Now, going back to the curved space metric ansatz%
\begin{equation}
ds^{2}=-f\left( r\right) dt^{2}+\frac{dr^{2}}{f\left( r\right) }+r^{2}\left(
d\theta ^{2}+\sin ^{2}\theta d\varphi ^{2}\right)
\end{equation}%
one obtains for $\beta =0,$ the exact solution from (14) as 
\begin{equation}
f\left( r\right) =1-\frac{2m}{r}-\frac{\Lambda }{3}r^{2}-\frac{P\alpha }{%
\sqrt{2}},
\end{equation}%
with $E\left( r\right) =0.$ Such a metric represents a global monopole \cite%
{8} with a deficit angle which is valid only for $P\neq 0.$ We note that
this represents a non-asymptotically flat black hole with mass, cosmological
constant and global monopole charge.

For the case of pure electric field let us consider, now, in (3) $P=0$ and
due to the sign problem we revise our square root term as $\sqrt{-F_{\mu \nu
}F^{\mu \nu }}$ in the action. Further, to remove the ambiguity in
arbitrariness of $E(r)$ from the Maxwell equation (4) we require that the
spacetime has constant scalar curvature. This restricts our $E(r)$ only to
be a constant. This yields with reference to the metric ansatz (2), as a
result of the Maxwell equation, for the choice $N(r)=1$ that one obtains $%
R(r)=r_{0}=$constant and $E(r)=E_{0}=$constant.

The $tt$ and $rr$ components of the Einstein's equations yield

\begin{equation}
\Lambda =\frac{1}{r_{0}^{2}}\text{, }
\end{equation}%
so that the solution for $f(r)$ takes the form 
\begin{equation}
f=-\left( \Lambda +\frac{\alpha E_{0}}{\sqrt{2}}\right) r^{2}+C_{1}r+C_{2}
\end{equation}%
where $C_{1}$ and $C_{2}$ are constants of integration. With this $f(r)$ the
line element reads%
\begin{equation}
ds^{2}=-f\left( r\right) dt^{2}+\frac{dr^{2}}{f\left( r\right) }%
+r_{0}^{2}\left( d\theta ^{2}+\sin ^{2}\theta d\varphi ^{2}\right)
\end{equation}%
in which the electric field ($E_{0}$) and cosmological constant ($\Lambda =%
\frac{1}{r_{0}^{2}}$) are both essential parameters. By setting $%
E_{0}=0=C_{1}$, it reduces to the Nariai \cite{9} line element. For this
reason (27) is known for $E_{0}\neq 0$ to be the Nariai-Bertotti-Robinson
(NBR)-type \cite{9} line element. Let us add that since our case is a NED,
rather than the linear Maxwell theory our solution shows minor digression
from the standard NBR spacetime \cite{9}. Due to this fact we prefer to
label it simply as NBR-type. In the sequel we consider radial geodesics for
both $P\neq 0$ and $P=0$.

\subsection{Absence of linear potential for $P\neq 0,$ $\protect\beta \neq 0$%
}

We study the radial geodesics of a charged particle with electric charge $%
q_{0}$ and unit mass ($m=1$) for simplicity in the spacetime (2) (for $%
N\left( r\right) =1$ and $R(r)=r$). For the radial geodesics we set $\theta
=\theta _{0}=$constant and $\varphi =\varphi _{0}=$constant, so that the
particle Lagrangian is given by (a 'dot' in the sequel stands for derivative
with respect to the proper distance $s$)

\begin{equation}
L=\frac{1}{2}\left( -f\ \dot{t}^{2}+\frac{1}{f}\dot{r}^{2}\right) +q_{_{0}}P%
\dot{t}\int \frac{P}{r^{2}\sqrt{1+\frac{r^{4}}{\beta ^{2}}}}dy.
\end{equation}%
Herein a constant of motion is given by%
\begin{equation}
\frac{\partial L}{\partial \dot{t}}=-\mathcal{E}
\end{equation}%
where $\mathcal{E}$ represents the energy of the particle. The geodesic
equation reads%
\begin{equation}
f\ \dot{t}=q_{0}P\int \frac{P}{r^{2}\sqrt{1+\frac{r^{4}}{\beta ^{2}}}}-%
\mathcal{E}
\end{equation}%
with geodesic condition 
\begin{equation}
\dot{r}^{2}+f=\left( f\ \dot{t}\right) ^{2}.
\end{equation}%
A substitution yields the equation of motion for the particle 
\begin{equation}
\dot{r}^{2}+V_{eff}=\mathcal{E}^{2}
\end{equation}%
where%
\begin{equation}
V_{eff}=1-\frac{2m}{r}-\frac{\Lambda }{3}r^{2}-\frac{P\alpha }{\sqrt{2}r}%
\int \sqrt{1+\frac{\beta ^{2}}{r^{4}}}dr-\left( q_{0}P\right) ^{2}\left(
\int \frac{dr}{r^{2}\sqrt{1+\frac{r^{4}}{\beta ^{2}}}}\right) ^{2}+2\mathcal{%
E}q_{0}P\int \frac{dr}{r^{2}\sqrt{1+\frac{r^{4}}{\beta ^{2}}}}.
\end{equation}%
Once more we expand this potential to get for small $r$%
\begin{equation}
V_{eff}\left( r_{small}\right) =1-\frac{2m+2\mathcal{E}q_{0}P}{r}+\frac{%
P\alpha \beta \sqrt{2}-2q_{0}^{2}P^{2}}{2r^{2}}-\left( \frac{P\alpha \sqrt{2}%
}{12\beta }+\frac{\Lambda }{3}+\frac{q_{0}^{2}P^{2}}{3\beta ^{2}}\right)
r^{2}-\frac{\mathcal{E}q_{0}P}{3\beta ^{2}}r^{3}+\mathcal{O}\left(
r^{6}\right) 
\end{equation}%
and for large $r$

\begin{equation}
V_{eff}\left( r_{large}\right) =1-\frac{1}{2}P\alpha \sqrt{2}-\frac{2m}{r}-%
\frac{\Lambda }{3}r^{2}-\frac{2\mathcal{E}q_{0}P\beta }{3r^{3}}+\frac{%
P\alpha \sqrt{2}\beta ^{2}}{12r^{4}}+\mathcal{O}\left( \frac{1}{r^{6}}%
\right) ,
\end{equation}%
where both manifestly show the absence of a linear ($\sim r$) term in the
effective potential $V_{eff}.$ Our expansions, however, cover only the
asymptotic regions for small / large $r$ values. For a general proof
arbitrary $r$ should be accounted which can be expressed in terms of
elliptic functions. 

\subsection{Linear potential for $P=0$ and $E=E_{0}=$constant}

With the constant electric field $E_{0}$ now we have $V(r)=-E_{0}r,$ up to a
disposable constant. The Lagrangian of a charged particle in the spacetime
(27) (with charge $q_{0}$ and unit mass) is given by 
\begin{equation}
L=\frac{1}{2}\left( -f\ \dot{t}^{2}+\frac{1}{f}\dot{r}^{2}\right)
+q_{0}E_{0}r\dot{t}.
\end{equation}%
For simplicity we set $r_{0}=q_{0}=1$ and $\alpha =2\sqrt{2}$ so that the
geodesic motion takes the form%
\begin{equation}
\left( \frac{dr}{ds}\right) ^{2}=\mathcal{E}^{2}+Ar^{2}+Br-C_{2}.
\end{equation}%
Here $\mathcal{E}>0$ is the conserved energy and the constants $A$ and $B$
are abbreviated as%
\begin{equation}
A=\left( 1+E_{0}\right) ^{2},\text{ \ \ }B=-C_{1}+2E_{0}\mathcal{E}.
\end{equation}%
We set now $A=0=C_{1},C_{2}=-1$, so that (37) integrates with the effective
linear potential $V_{eff}=2\mathcal{E}r$ to yield 
\begin{equation}
\sqrt{\mathcal{E}^{2}-2\mathcal{E}r+1}=\pm \mathcal{E}\left( s_{0}-s\right) 
\end{equation}%
in which $s_{0}$ is an integration constant. Clearly the potential is
confined by $0<r<\frac{\mathcal{E}^{2}+1}{2\mathcal{E}}$ and the underlying
geometry is a NBR spacetime transformable (for $r=\cosh \chi $) to the line
element 
\begin{equation}
ds^{2}=-\sinh ^{2}\chi dt^{2}+d\chi ^{2}+d\theta ^{2}+\sin ^{2}\theta
d\varphi ^{2}
\end{equation}%
with electric field $E_{0}=-1$ and cosmological constant $\Lambda =1.$

In conclusion, the square root Lagrangian $\sqrt{F_{\mu \nu }F^{\mu \nu }}$
(or $\sqrt{-F_{\mu \nu }F^{\mu \nu }}$ for the pure electric case) with a
cosmological constant in the absence of the standard Lagrangian $F_{\mu \nu
}F^{\mu \nu }$ admits solution with a uniform electric field (and zero
magnetic charge) which provides a linear potential believed to be effective
in confinement \cite{6}. The Coulomb part of the 'Cornell potential' \cite%
{10} will naturally be absent in our formalism. We have shown also that
magnetic charge ($P\neq 0$) acts against confinement. The spacetime in which
the square root Maxwell Lagrangian yields confinement happens to be the
constant curvature NBR spacetime even in this form of the square root NED.
In such a spacetime we have both electric field, cosmological constant and
the freedom of choice of one in terms of the other renders a linear
potential in the effective potential $V_{eff}$ possible.

\textbf{Acknowledgment:} \textit{The authors would like to thank the
anonymous reviewer for his / her valuable and helpful comments.}

\end{document}